\documentstyle[12pt] {article}
\def\be{\begin{equation}}
\def\ee{\end{equation}}
\def\bea{\begin{eqnarray}}
\def\eea{\end{eqnarray}}
\def\pd{\partial}
\begin{document}
\title{Brane-wave Duality}
\author{Linda Baker$\footnote{e-mail: l.m.baker@durham.ac.uk}$\,  and 
David Fairlie$\footnote{e-mail: david.fairlie@durham.ac.uk}$\\
\\
Department of Mathematical Sciences,\\
         Science Laboratories,\\
         University of Durham,\\
         Durham, DH1 3LE, England}
\maketitle
\begin{abstract}
 The quantum mechanical transition between a free particle Lagrangian  and the Klein Gordon field description of a free particle (particle wave duality) is conjectured to extend to
 an analogous construction of relativistically invariant wave equations 
 associated with strings and branes, which we propose to call brane-wave 
 duality. 
Electromagnetic interactions in the two systems are discussed.
It is emphasised that all integrable free field theories, including those of 
Dirac-Born-Infeld type, are associated with  Lagrangians equivalent to 
divergences on the space of solutions of the equations of motion.
 
\end{abstract}
 
\section{Introduction}
In standard textbooks on Quantum Mechanics, the description of free  particle motion by the classical point particle Lagrangian
\be
{\cal L}_1\,=\,\sqrt{\sum\left(\frac{\pd X^\mu}{\pd\tau}\right)^2}\label{one}
\ee
goes over in terms of a  field theory, to that given by the Lagrangian of a Klein Gordon field;
\be
{ L}_2\,=\,\ \frac{1}{2}{\sum\left(\frac{\pd \phi}{\pd x_\mu}\right)^2}\label{KG3}
\ee
(for a massless field). This may be cited as an example of particle-wave duality. To distinguish between Lagrangians which involve
 square roots and those which do no we use a cursive notation for the former and
a capital for the latter. 
Is there a similar alternative description of strings and branes? One goal of this article is to suggest that there is a natural extension. The idea is that
corresponding to each direction in the world-volume there should be associated a field. For example
the strings described by the Nambu-Goto String Lagrangian
\be
{\cal L}_3\,=\,\sqrt{\sum\left[\left(\frac{\pd X^\mu}{\pd\sigma}\frac{\pd X^\mu}
{\pd\tau}\right)^2-\left(\frac{\pd X^\mu}{\pd\sigma}\right)^2\left(\frac{\pd X^\nu}{\pd\tau}\right)^2\right]}\label{three}
\ee
should also admit a description in terms of two fields with Lagrangian which is
some power of the following
\bea
{ L}_4&=&\sum\left[\left(\frac{\pd \phi}{\pd x_\mu}\frac{\pd \psi}{\pd x_\mu}\right)^2-
\left(\frac{\pd \phi}{\pd x_\mu}\right)^2\left(\frac{\pd \psi}{\pd x_\nu}\right)^2\right]\nonumber\\
&=& -\frac{1}{2}\sum_{\mu,\ \nu}\left (\frac{\pd \phi}{\pd x_\mu}\frac{\pd \psi}
{\pd x_\nu}-\frac{\pd \phi}{\pd x_\nu}\frac{\pd \psi}{\pd x_\mu}\right)^2
\label{four}
\eea
Here $\phi(x_\mu),\ \psi(x_\mu)$ are two fields and $\mu,\ \nu$ range over the dimensionality of space time. Likewise,  a simple brane Lagrangian, $\displaystyle{\sqrt{\det\left|\frac{\pd X^\mu}{\pd\sigma_i}\frac{\pd X^\mu}{\pd\sigma_j}\right|}}$ may be conjectured to be equivalent to a
field theory with as many components as there are world-volume co-ordinates  with Lagrangian a power of
\be
{ L}\,=\,{\det\left|\frac{\pd \phi^i}{\pd x_\mu}\frac{\pd \phi^j}{\pd x_\mu}\right|}\,=\, \left(\frac{n!(d-n)!}{d!}\right)\sum\left(\frac{\pd\{\phi^1,\ \phi^2,\dots,\phi^n\}}{\pd\{x_{\mu_1},x_{\mu_2}\dots x_{\mu_n}\}}\right)^2.\label{field}
\ee
Here the sum is over all permutations of the squares of Jacobians of the $n$ fields  with respect to  selections of $n$ (the dimension of the world volume) out of the $d$ co-ordinates $x_\mu$ of space-time.
This higher dimensional analogue of particle-wave duality
we might facetiously describe as brane-wave duality. Similar ideas have been advanced before by Hosotani \cite{hos1}\cite{hos2} and Morris \cite {morris1}\cite{morris2}. What they do is to take the Lagrangian
for a string in $d$-dimensions and relate it to a Lagrangian for $d-2$ fields in  $d$ dimensions. This amounts to a change of variables in their case. They show that the classical equations of motion are equivalent under interchange of dependent and independent variables. However, we should like to advocate that since the quantum mechanical equation which describes a free particle is the Klein Gordon equation, for only one field, independent of the dimension of the embedding space, it would seem that in the case of $n$ branes the field theories describing them quantum mechanically should depend upon only $n$,  rather than $d-n$ fields. 
  We shall argue however, that it seems more attractive to take a  square root
as the correct power, as this, as we shall demonstrate, guarantees covariance
of the equations of motion; the counterpart of the reparametrisation invariance of the original Dirac-Born-Infeld brane Lagrangians.

The two properties found for the Nambu-Goto Lagrangian $({\cal L}_3)$, namely  that it transforms under reparametisation of the independent variables with a factor which is the  Jacobian of the transformation and that
${\cal L}_3$ vanishes or is constant on the space of solutions of the equations
of motion of the Schild Lagrangian $({\cal L}_3)^2$,\cite{schild}\cite{eguchi},
have recently been shown to persist for Lagrangians of the Dirac-Born-Infeld type 
\cite{born}\cite{dirac}\cite{tseytlin} ,
\be 
{\cal L}\,=\,\sqrt{\det{| g_{ij}+F_{ij} |}}\,=\, \sqrt{\det\left| \frac{\partial X^\mu}{\partial x_i}\frac{\partial X_\mu}{\partial x_j}\,+\,\left( \frac{\partial A_j}{\partial x_i}\,-\,\frac{\partial A_i}{\partial x_j}\right)\right|}\ .\label{born}
\ee
even in the presence of electromagnetic fields \cite{dbf}. Analogues of both 
these properties may be deduced for the wave field Lagrangians introduced in 
this paper (\ref{field}). We see the first of these as being common to all 
theories describing free objects, and devote the next section to a catalogue 
of 
examples. These may in certain cases be extended to include supersymmetry. 
 We then demonstrate how the diffeomorphism invariance of the brane Lagrangians 
translates into covariance of solutions for the wave fields with square root 
Lagrangians. We then analyse in some detail the case of the point particle and 
the string.
Some special cases of these Lagrangians are shown to be completely integrable in implicit 
form  and finally we
make some speculative remarks concerning the coupling of gauge theories 
to the wave fields.

\section{ Lagrangians for free fields}

In fact we may note that a generic property of free fields is that their Lagrangians vanish, or are divergences on the space of solutions of the equations of motion which vanish, with vanishing first derivatives on the boundary. This is trivial for ${ L}_2$, the Klein Gordon Lagrangian, which may be rewritten after partial integration as
\be
{ L}_2\,=\,\ -\frac{1}{2}\sum \phi\frac{\pd^2 \phi}{\pd x_\mu^2}.\label{KG2}
\ee
Likewise the Dirac Lagrangian, $ i\bar\psi\gamma_\mu\pd_\mu\psi-m\bar\psi\psi$
which is already in null form, that for the Maxwell theory and pure gravity all
vanish on the space of solutions of the equations of motion. It is not true of non abelian gauge theory, except for the self dual sector, where, by definition
\be
 F_{\mu\nu} F^{\mu\nu}\,=\, \frac{1}{2}\epsilon_{\mu\nu\rho\sigma} F^{\mu\nu}F^{\rho\sigma},
\nonumber
\ee
and the right hand side is a divergence.
This principle extends to the wave Lagrangians of Born-Infeld type   in a straightforward manner.
 Here, under field  redefinitions $\phi\rightarrow \Phi(\phi,\psi),\  \psi\rightarrow \Psi(\phi,\psi)$ the Lagrangian ${\cal L}_4\,=\, \sqrt{L_4}$ scales with a conformal factor which is the Jacobian of the transformation. Also the quantity
\be 
\phi\frac{\pd}{\pd x_\mu}\left(\frac{\pd{L}_4}{\pd \frac{\pd \phi}{\pd x_\mu}}\right) \label{eqmot1}
\ee
considered as a Lagrangian, reproduces the same equation of motion as does ${L}_4$ and is zero on the space of solutions. It thus differs from ${L}_4$ by a divergence.
In a similar fashion
\be 
\psi\frac{\pd}{\pd x_\mu}\left(\frac{\pd{L}_4}{\pd \frac{\pd \phi}{\pd x_\mu}}\right) \label{eqmot2}
\ee
is itself a divergence. 
Similar remarks, mutatis mutandis, can be made about the same constructions with
$\phi\leftrightarrow\psi$.
These statements are easy to prove and simply depend upon the determinantal 
nature of ${L}_4$. The key result, which obtains also in general dimension for (\ref{field}) is that
\be 
\sum_\mu\frac{\pd\phi^i}{\pd x_\mu}\frac{\pd{ L}}{\pd \frac{\pd \phi^j}{\pd x_\mu}} \,=\, 2\delta^i_j{ L}.\label{key}
\ee
Consider
\be 
\phi\frac{\pd}{\pd x_\mu}\left(\frac{\pd{L}_4}{\pd \frac{\pd \phi}{\pd x_\mu}}\right)= \frac{\pd}{\pd x_\mu}\left(\phi\frac{\pd{L}_4}{\pd \frac{\pd \phi}{\pd x_\mu}}\right)-\frac{\pd\phi}{\pd x_\mu}\left(\frac{\pd{L}_4}{\pd \frac{\pd \phi}{\pd x_\mu}}\right). \label{eqmot3}
\ee
The first term on the right hand side is a  divergence, the second is simply
 $-2{L}_4$, as a consequence of the determinantal properties of ${L}_4$. Thus the left hand side of (\ref{eqmot3}) serves as an equivalent Lagrangian, giving the same equations of motion as the original one. In the 
case of  (\ref{eqmot2}) the corresponding term left over after extracting the divergence is zero, again as a consequence of determinantal properties.
Obviously this argument extends to dual brane Lagrangians of any dimension. This property suggests that these Lagrangians have a pseudo-topological aspect. If they were equivalent to divergences without any further constraint, they would be fully topological; but here the equivalence works only for solutions of the equations of motion.

\section{Equations of Motion of Born-Infeld Type}
For the classical point particle Lagrangian (\ref{one}) then the equations of motion can be written as
\be
\frac{\pd^2 X^{\mu}}{\pd \tau^2} \frac{\pd X^{\nu}}{\pd \tau}-\frac{\pd^2 X^{\nu}}{\pd \tau^2} \frac{\pd X^{\mu}}{\pd \tau}=0.\label{class1}
\ee
For $d$ dimensions then it is easy to verify that there are $d-1$ independent equations of motion.
Now consider the string case in d=3 dimensions. The Nambu-Goto Lagrangian ${\cal L}_3$ gives the single equation of motion
\be
\left( \begin{array}{ccc}
\hat{J}_1 & \hat{J}_2 & \hat{J}_3
\end{array} \right) 
\left( \begin{array}{ccc}
X^1_{\sigma\sigma} & X^1_{\sigma\tau} & X^1_{\tau\tau}\\
X^2_{\sigma\sigma} & X^2_{\sigma\tau} & X^2_{\tau\tau}\\
X^3_{\sigma\sigma} & X^3_{\sigma\tau} & X^3_{\tau\tau} 
\end{array} \right)  
\left(  \begin{array}{c}
(X^1_\tau)^2+(X^2_\tau)^2+(X^3_\tau)^2\\
-2(X^1_\sigma X^1_\tau+X^2_\sigma X^2_\tau+X^3_\sigma X^3_\tau)\\
(X^1_\sigma)^2+(X^2_\sigma)^2+(X^3_\sigma)^2
\end{array} \right) =0\label{class2}
\ee
where
\be
X^\mu_{ij}=\frac{\pd^2 X^{\mu}}{\pd \sigma^i \pd \sigma^j},\qquad  X^\mu_i=\frac{\pd X^\mu}{\pd \sigma^i}, \qquad \textrm{and}\quad \sigma^i=(\sigma, \tau), \label{class3}
\ee
\be
\hat{J}_\rho=\epsilon_{\rho\mu\nu} X^\mu_\sigma X^\nu_\tau =\frac{1}{2} \epsilon_{\rho\mu\nu}
\left| \begin{array}{cc}
X^\mu_\sigma  & X^\nu_\sigma\\
X^\mu_\tau & X^\nu_\tau
\end{array} \right|\label{class4}
\ee
In general, a typical equation of motion, of which only $d-2$ are independent can be written in the following form:
\be
\hat{J}_\nu X^\nu_{ij} (L^{-1})_{ij} = 0,\label{class5}
\ee
where L is the matrix with components $[L]_{ij}=\frac{\pd X^\mu}{\pd\sigma^i}\frac{\pd X^\mu}{\pd \sigma^j}$ and $\nu$  is chosen from three of the values $\nu_1,\ \nu_2,\ \nu_3$  of the index $\mu$ which runs over $1\dots d$. $\hat J_{\nu_1}$ denotes the Jacobian $\displaystyle{\frac{\pd( X^{\nu_2}, X^{\nu_3})}{\pd(\ \sigma_1,\ \sigma_2\ )}}$, omitting $X^{\nu_1}$ etc.
 This can be extended to strings in $d$ dimensions and to branes. The only essential difference is that in the typical equation of motion, $\nu$ is now an arbitrary choice of $n+1$ values and $\hat J_\nu$ is now a Jacobian of a subset of $n$ of those  variables
$x^\nu$, with respect to the $n$ world sheet co-ordinates
$\sigma_j$. 

Computer calculations show that for a the Nambu-Goto Lagrangian there are $d-2$ independent equations of motion. In general, an object (particle/string/brane)  which sweeps out an $n$-dimensional world volume in $d$-dimensional space-time has only $d-n$ independent equations of motion. The basic reason for this is that in the case $d\,=\,n$ the Lagrangian is a divergence, so all the equations
of motion vanish.
\section{Equations of Motion of Inverse Type}
For the Lagrangians of inverse type, i.e. ${\cal L}_2\,=\, \sqrt{L_2},\  {\cal L}_4\,=\, \sqrt{L_4}$ etc. there is just one equation for  ${\cal L}_2$ and two
for  ${L}_4$ and so on, irrespective of the dimension of the total space; however these equations fall into sums of equations appropriate to the minimal 
dimension for a non trivial embedding. What this means is that for  ${\cal L}_2$
in 2 dimensions, the minimal case the equation is the well known Bateman equation \cite{gov}
\be
\left(\frac{\pd \phi}{\pd x_1}\right)^2\frac{\pd^2 \phi}{\pd x_2^2}+
\left(\frac{\pd \phi}{\pd x_2}\right)^2\frac{\pd^2 \phi}{\pd x_1^2}-
2\left(\frac{\pd \phi}{\pd x_1}\right)\left(\frac{\pd \phi}{\pd x_2}\right)\frac{\pd^2 \phi}
{\pd x_1\pd x_2}\,=\,0,\label{batman}
\ee
 to be discussed at greater length in section 5, while in 3 dimensions the equation 
is the sum of three Bateman equations, corresponding to the three ways of selecting two co-ordinates out of three. A particular class of solutions to these equations can be found by simultaneously setting these three Bateman equations to zero. It is also noteworthy that these three equations also result from the transformation of the three equations of the form (\ref{class1}) by exchanging the r\^oles of dependent and independent variables. 
Much the same happens for  ${L}_4$. Here the minimal dimension is 3. The 
two equations of motion in $d$ dimensions fall into the sum of $d\choose 3$ copies
of the minimal equations, whose solution is discussed in section 5. \section{Covariance}
This chapter deals with special properties of Lagrangian densities of the previous form, but where the square root has been taken, as in the standard Born-Infeld. Here the feature of general covariance plays an important r\^ole.
This works for arbitrary dimension; under the transformation of the Lagrangian ${\cal L}$ of square root type under the field redefinition
\be
\phi^i\to \Phi^i(\phi^1, \phi^2,\dots,\phi^n), \label{mapsto}
\ee
${\cal L}$ acquires a factor which is the Jacobian of the transformation
and the equations of motion are unaffected on account of (\ref{key}) since they are
given by
\bea
&&\frac{\pd J}{\pd \phi^i}{\cal L}-\frac{\pd }{\pd x_\mu}\left(J\frac{\pd{\cal L}}{\pd \frac{\pd \phi}{\pd x_\mu}}\right)\nonumber\\
&=&\frac{\pd J}{\pd \phi^i}{\cal L}-\frac{\pd J}{\pd \phi^j}\frac{\pd\phi^j }{\pd x_\mu}\frac{\pd{\cal L}}{\pd \frac{\pd \phi}{\pd x_\mu}}-J\frac{\pd }{\pd x_\mu}\left(\frac{\pd{\cal L}}{\pd\frac{\pd \phi}{\pd x_\mu}}\right)\label{neweq}\\
&=&-J\frac{\pd }{\pd 
x_\mu}
\left(\frac{\pd{\cal L}}{\pd\frac{\pd \phi}{\pd x_\mu}}\right)\,=\,0,
\nonumber
\eea
as the first two terms cancel on account of (\ref{key}). 
Thus the square root of the determinantal form is a generally
covariant Lagrangian, which means that any function of a 
solution remains a solution of the equations of motion.  In fact the equations of motion arising from these Lagrangians,
in the case where the number of co-ordinates $x_\mu$ exceeds 
that of the number $n$ of wave fields by one are a 
generalisation of the Bateman equation\footnote{for a 
generalisation in a somewhat different direction see \cite{gov}}
and are expected to be completely integrable, since this is the case for $n=1$ and $n=2$, as we shall demonstrate in
the next section. 
\subsection{Integrability in special cases}
Consider the Lagrangian
\be
{\cal L}_{\rm two}\,=\,\ \sqrt{\sum\left(\frac{\pd \phi}{\pd x_\mu}\right)^2}\label{KG}
\ee
in the case of 2 dimensions ($\mu =1,\ 2$).
This action is fully integrable as the equation of motion is just the 
 Bateman equation (\ref{batman})

This equation has the general solution;
\be 
F(\phi)x_1+G(\phi)x_2\,=\,c\,=\ \ {\rm constant}\label{soln}
\ee
where $F,\ G$ are two arbitrary functions. It is clearly covariant; if $\phi$
is a solution so is any function of $\phi$.
It is equivalent to a Monge nonlinear wave equation
\be
\frac{\pd u}{\pd x_1}\, =\,u\frac{\pd u}{\pd x_2}\label{monge}
\ee
where 
 \be
 u\, =\,\frac{ \frac{\pd \phi}{\pd x_1}}{ \frac{\pd \phi}{\pd x_2}}.
 \label{monge2}
\ee 
 What is the situation with the next Lagrangian with 2 fields in 3 dimensions
($\mu,\nu =1,\ 2,\ 3$),
\be
{\cal L}_{\rm three}=\sqrt{\sum\left[\left(\frac{\pd \phi}{\pd x_\mu}\frac{\pd \psi}
{\pd x_\mu}\right)^2-\left(\frac{\pd \phi}{\pd x_\mu}\right)^2\left(\frac{\pd \psi}
{\pd x_\nu}\right)^2
\right]}\ ?
\label{four1}
\ee
We know that the equations of motion are covariant; therefore we expect that 
they are expressible in first order form in terms of two ratios of the Jacobians
\bea
 u&=&\frac{ \phi_{x_1}\psi_{ x_2}- \phi_{x_2} \psi_{x_1}}
 { \phi_{x_2}\psi_{ x_3}- \phi_{x_3} \psi_{x_2}}
\nonumber\\
v&=& \frac{ \phi_{x_3}\psi_{ x_1}- \phi_{x_1} \psi_{x_3}}
 { \phi_{x_2}\psi_{ x_3}- \phi_{x_3} \psi_{x_2}}\label{newdef}
\eea
where $\phi_{x_\mu}$ denotes the partial derivative 
$\frac{\pd \phi}{\pd x_\mu}$.
The Lagrangian in the first case takes the form
\be
{\cal L}_{\rm two} =\phi_{x_2}\sqrt{(1+u^2)}\label{case1}
\ee
Working out the equation of motion gives 
\be
\frac{\pd}{\pd x_2}\frac{1}{\sqrt{1+u^2}}+
\frac{\pd}{\pd x_1}\frac{u}{\sqrt{1+u^2}}\,=\,0.\label{ex}
\ee
This equation is equivalent to (\ref{monge}).
The remarkable feature of this Lagrangian is that any differentiable function 
$f(u)$  instead of ${\sqrt{1+u^2}}$ will give the same equations of motion 
\cite{gov}!
This generalises;
In the second case
\be
{\cal L}_{\rm three} =\left(\phi_{x_2}\psi_{ x_3}- \phi_{x_3} \psi_{x_2}\right)
\sqrt{(1+u^2+v^2)}\label{case2}
\ee
and the two independent equations of motion are
\bea
\frac{\pd}{\pd x_2}\frac{1}{\sqrt{1+u^2+v^2}}-
\frac{\pd}{\pd x_1}\frac{v}{\sqrt{1+u^2+v^2}}&=&0.\nonumber\\
\frac{\pd}{\pd x_3}\frac{1}{\sqrt{1+u^2+v^2}}-
\frac{\pd}{\pd x_1}\frac{u}{\sqrt{1+u^2+v^2}}&=&0.\label{exq}
\eea
$\sqrt{(1+u^2+v^2)}$ may be replaced by any arbitrary differentiable 
function of two variables, $f(u,v)$ and an equivalent pair of equations of 
motion is  as follows;
\bea
\frac{\pd u}{\pd x_1}&=&u\frac{\pd u}{\pd x_3}+v\frac{\pd u}{\pd x_2}
\nonumber\\
\frac{\pd v}{\pd x_1}&=&u\frac{\pd v}{\pd x_3}+v\frac{\pd v}{\pd x_2}
\label{eex}
\eea
These equations admit an implicit solution for $(u,\ v)$ given by solving the 
equations
\be
u\,=\,F(x_3+ux_1,\ x_2+vx_1),\quad\quad v\,=\,G(x_3+ux_1,\ x_2+vx_1)\label{solun}
\ee
where $F,\ G$ are two arbitrary functions of two variables. The general 
solution to the equations of motion is given by setting 
$u\,=\, U(\phi,\ \psi)$ and $v\,=\, V(\phi,\ \psi)$ where $U,\ V$ are two 
further arbitrary functions, and solving (\ref{eex}) for $\phi,\ \psi$.
This demonstrates the integrability of the equations of motion. The 
generalisation to $n$ fields in $n+1$ dimensions will be straightforward. It is 
anticipated that there will be an integrable generalisation to $n$ fields in $d$ dimensions along the lines of the Universal Field Equation \cite{gov}\cite{gov2}, but that
would take us too far afield and away from the spirit of the present 
investigation.
\section{Electromagnetic interactions}
The subject of electromagnetic interactions in the study of these new 
 Lagrangians is at this stage somewhat speculative. In the original theory, electromagnetic interactions are  implemented by the rather ad-hoc procedure of adding an
antisymmetric piece to  the induced metric;
$g_{ij}\rightarrow g_{ij} +  F_{ij}$
This is consistent with the picture of having gauge fields living on the
brane, and presumably confined to it by some Meissner type effect. However the natural way to couple electromagnetic fields to the dual theory is through 
a coupling to the conserved currents in the theory. This will ensure gauge invariance. It is easy to construct such conserved quantities, eg
\be
J^{ij}_\mu\,=\,\frac{\pd{\cal L}}{\pd\frac{\pd \phi^i}{\pd x_\mu}}\phi^j,\ \ \ \ i\,\neq\, j
\label{cons1}
\ee 
and  in the case $i=j$, the currents $J^{ii}-J^{jj}$ are also conserved.
There is however an embarasse de richesse here as these currents  carry two indices. Thus the natural coupling is to a  two index gauge field $A^{ij}_\mu$ transforming under $SO(n)$, i.e. the contribution
\be
{\cal L}_{\rm current}\, =\, \sum_{i,j} A^{ij}_{\mu}J^{ij}_\mu\label{couple}
\ee
represents the part of the Lagrangian coupling a non Abelian gauge field to
the fields $\phi^i$. This suggestion, though appealing, is rather unorthodox.
It is however gauge invariant up to a divergence, as in the gauge transformation
\be
A^{ij}_\mu\to O^{ik}A^{kl}_\mu O^{lj}+O^{ik}\frac{\pd}{\pd x_\mu}O^{kj}\label{gauge}
\ee
the rotation matrices may be removed by a linear transformation of the fields $\phi^i$ and the inhomogeneous term converted to an innoccuous divergence. 
If one wants instead to mimic the Dirac-Born-Infeld incorporation of electromagnetism,  
the simplest assumption would be to suppose that $A_\mu$ depends 
upon the co-ordinates $x_\nu$ only through their dependence upon 
$\phi^j$ with direction in a linear combination of the gradients 
of $\phi^j$. Then we may set
\be
A_\mu\, =\,\sum_j\frac{\pd\phi^j}{\pd x_\mu}{\cal A}^j(\phi^i).\label{ansatz}
\ee
Then a new Lagrangian which adds a term to the matrix components anti-symmetric in $(i,\ j)$, the antisymmetry being related to that of $F_{\mu\nu}$ takes the form with $L^{ij}\,=\,\frac{\pd \phi^i}
{\pd x_\mu}\frac{\pd \phi^j}{\pd x_\mu}$
\bea
{\cal L}'&=&\sqrt{\det\left|\frac{\pd \phi^i}{\pd x_\mu}\frac{\pd \phi^j}{\pd x_\mu}
+\frac{\pd}
{\pd\frac{\pd \phi^i}{\pd x_\mu}}\log({\cal L})
\frac{\pd}{\pd\frac{\pd \phi^j}{\pd x_\nu}}\log({\cal L})
\left(\frac{\pd A_\mu}{\pd x_\nu}-\frac{\pd A_\nu}{\pd x_\mu}\right)\right|}\nonumber\\
&=&\sqrt{\det\left|L^{ij}+(L^{-1})^{ip}(L^{-1})^{qj}\frac{\pd \phi^p}{\pd x_\mu}
\frac{\pd \phi^q}{\pd x_\nu}\left(\frac{\pd A_\mu}{\pd x_\nu}-
\frac{\pd A_\nu}{\pd x_\mu}\right)\right|}\nonumber\label{compare}\\
&=&\sqrt{\det\left|\frac{\pd \phi^i}{\pd x_\mu}\frac{\pd \phi^j}{\pd x_\mu}+
\left(\frac{\pd{\cal A}^i}{\pd \phi^j}-\frac{\pd {\cal A}^j}{\pd \phi^i}
\right)\right|}.\nonumber
\eea 
This manifestly gauge invariant suggestion is in the same spirit as the original Born-Infeld Lagrangian especially if the square root form is taken.   

An alternative suggestion to compare the Kalb-Ramond string interaction
\be
B_{\mu\nu}\frac{\pd X^\mu}{\pd\sigma}\frac{\pd X^\nu}{\pd\tau}
\ee seems less satisfactory, because on the dual brane side this would give
\be
B_{\mu\nu}\frac{\pd \phi}{\pd x_\mu}\frac{\pd \psi}{\pd x_\nu}
\ee
and this interaction, though gauge invariant is more like a Pauli term than 
one coming from minimal coupling.
\section{Conclusion}
While our answer might require some modification, we believe that we have posed
an interesting and important question; to find   Lagrangians for fields
which bear a similar relation to String and Brane Lagrangians as does the Klein Gordon to that of the classical point particle. A common characteristic of these Lagrangians, as indeed of all free theories, is that they are pseudo-topological, i.e. the Lagrangian is equivalent to a divergence on the space of solutions of the equations of motion. If the square root form is taken then the new Lagrangians are covariant, and the equations of motion are integrable and coincide in the case that the dimensions of space-time exceeds the that of the fields by one, with the Universal Field Equations proposed earlier \cite{gov}\cite{gov2}.  
 Some ideas for the introduction of gauge fields have been discussed.
\section*{Acknowledgement}
Linda Baker is grateful to EPSRC for a postgraduate research award.


\newpage
 \begin{thebibliography}{9}
\bibitem{hos1} Y. Hosotani, {\it Phys.Rev.Letters} {\bf 47} (1981)
399.
\bibitem{hos2}Y. Hosotani and R. Nakayama, `The Hamilton-Jacobi Equations
for Strings and Membranes. {\bf hep-th/9903193} (1999)
\bibitem{morris1} T.R. Morris, From First to Second Quantized String Theory,
{\it Phys. Lett.} {\bf 202B} (1988) 222.
\bibitem{morris2} D.L. Gee and T.R. Morris,  From First to Second Quantized String Theory. 2. The Dilaton and other Fields, {\it Nucl. Phys} {\bf B331} (1990) 675.\\
 D.L. Gee and T.R. Morris,  From First to Second Quantized String Theory. 3.
Gauge Fixing and Quantization, {\it Nucl. Phys} {\bf B331} (1990) 694.
\bibitem{schild} A. Schild, { Phys. Rev.} {\bf D16} (1977) 1722.\\
\bibitem{eguchi}T. Eguchi,  {\it Phys. Rev. Lett.} {\bf 44} (1980) 126.
\bibitem{born} M. Born and L. Infeld Proc. Roy. Soc. {\bf A144} (1934) 425.
\bibitem{dirac}P.A.M. Dirac, An extensible model of the electron  Proc. 
Roy. Soc. {\bf A268} (1962) 57-67.
\bibitem{tseytlin}A.A. Tseytlin, Born-Infeld action, supersymmetry and string 
theory, {\bf hep-th/9908105} (1999) to appear in the Yuri Golfand memorial 
volume.
\bibitem{dbf} D.B. Fairlie,  Dirac-Born-Infeld Equations, {\it Phys .Lett.} {\bf B456}, (1999) 141-146.
\bibitem{suzy} I.L. Buchbinder and S.M. Kuzenko, {\it Ideas and Methods of Supersymmetry and Supergravity} I.O.P. Publishing, Ltd. (1995) (ISBN 0 75030 258 5).
\bibitem{gov} D.B. Fairlie, J. Govaerts and A. Morozov, Universal 
Field Equations
with Covariant Solutions, {\it Nuclear Physics B 373} (1992) 214-232.
\bibitem{gov2} D.B. Fairlie and J. Govaerts, Euler Hierarchies and Universal Equations,
{\it Journal of Mathematical Physics}.{\bf 33}\ (1992)\ 3543-3566. 
\bibitem{rev} D.B.Fairlie Integrable Systems in Higher Dimensions 
{\it Quantum Field Theory, Integrable Models and Beyond} 
Editors. T. Inami and R. Sasaki {\it Progress of Theoretical Physics Supplement}
 {\bf 118} (1995) 309-327.
\end{thebibliography}
\end{document}